 \documentclass[manuscript]{emulateapj}

 \usepackage{apjfonts}

\slugcomment{to appear in the Astrophysical Journal, Letters}

\shorttitle{Progenitors of Type Ia Supernovae}
\shortauthors{Hachisu et al.}

\begin{document}

\title{Final Fates of Rotating White Dwarfs and Their Companions
in the Single Degenerate Model of Type I\lowercase{a} Supernovae
}


\author{Izumi Hachisu}
\affil{Department of Earth Science and Astronomy,
College of Arts and Sciences, The University of Tokyo,
Komaba 3-8-1, Meguro-ku, Tokyo 153-8902, Japan}
\email{hachisu@ea.c.u-tokyo.ac.jp}

\author{Mariko Kato}
\affil{Department of Astronomy, Keio University,
Hiyoshi 4-1-1, Kouhoku-ku, Yokohama 223-8521, Japan}
\email{mariko@educ.cc.keio.ac.jp}

\and

\author{Ken'ichi Nomoto}
\affil{Kavli Institute for the Physics and Mathematics of the Universe,
The University of Tokyo, Kashiwanoha 5-1-5, Kashiwa, Chiba 277-8583, Japan}
\email{nomoto@astron.s.u-tokyo.ac.jp}



\begin{abstract}
Taking into account the rotation of mass-accreting white dwarfs (WDs)
whose masses exceed the Chandrasekhar mass, we extend our new single
degenerate model for the progenitors of Type Ia supernovae (SNe Ia),
accounting for two types of binary systems, those with a main sequence
companion and those with a red-giant (RG) companion.
  We present a mass distribution of WDs
exploding as SNe~Ia, where the WD mass ranges from 1.38 to
$2.3~M_\sun$.  These progenitor models are assigned to various types
of SNe~Ia. A lower mass range of WDs ($1.38~M_\sun<M_{\rm 
WD}\lesssim1.5~M_\sun$), which are supported by rigid rotation,
correspond to normal SNe~Ia.  A variety of spin-down time may lead
to a variation of brightness.  A higher mass range of WDs ($M_{\rm 
WD}\gtrsim1.5~M_\sun$), which are supported by differential rotation,
correspond to brighter SNe~Ia such as SN~1991T.  In this case, 
a variety of the WD mass may lead to a variation of brightness.
We also show the evolutionary states of the companion stars
at SN Ia explosions and pose constraints on the unseen companions.
In the WD+RG systems, in particular, most of the RG companions have
evolved to helium/carbon-oxygen WDs in the spin-down phase
before the SN~Ia explosions.  In such a case, we do not expect any
prominent signature of the companion immediately before and after
the explosion.  We also compare our new models with the recent stringent
constraints on the unseen progenitors of SNe~Ia such as SN~2011fe.
\end{abstract}




\section{Introduction}
\label{intro}
Type Ia supernovae (SNe~Ia) play important roles in astrophysics
as a standard candle for measuring cosmological distances and
as main production sites of iron group elements.
It is commonly agreed that the exploding star is
a mass-accreting carbon-oxygen (C+O) white dwarf (WD).
However, it is not clarified yet whether
the WD accretes H/He-rich matter from
its binary companion [single degenerate (SD) scenario], or two C+O WDs
merge [double degenerate (DD) scenario] \citep[e.g.,][]{hil00, nom00}.

Observations have provided the following constraints on the nature
of companion stars.  Some evidences support the SD model,
such as the presence of circumstellar matter (CSM) 
\citep{pat07, ste11, fol12} and detections of hydrogen 
in the circumstellar-interaction type
SNe~(Ia/IIn) like SN~2002ic \citep{ham03} and PTF11kx \citep{dil12}.
On the other hand, there has been no direct indication of
the presence of companions, e.g., 
(1) the lack of companion stars in the images of
SN~2011fe \citep{liw11}, some SN Ia remnants (SNRs) \citep{sch12},
SN 1572 (Tycho) \citep{ker09} and SN 1006 \citep{ker12},
(2) the lack of ultraviolet (UV) excesses
 of early-time light curves \citep{kas10}, and
(3) the lack of hydrogen features in the spectra \citep{leo07}.
Both (2) and (3) are expected from the collision between ejecta
and a companion.

In particular, detailed observations of SN~2011fe in M101 require
stringent constraints on the progenitor, i.e., \citet{liw11} excluded
the presence of a red-giant (RG), a helium star,
or a main-sequence (MS) star of $\gtrsim3.5~M_\sun$. 
\citet{bro12} further excluded
a solar mass MS companion from an early UV observation with
{\it Swift} because of no signature of shock interaction between ejecta
and a companion.

The tightest constraints come from no signature of shock interaction
with the companion.  However, if the binary separation, $a$, is much
larger than the companion radius, $R$, i.e., $a\gg R$, the solid angle
subtended by the companion would be much smaller, and so would be the effect
of shock interaction.  In their spin-down scenario, \citet{jus11} and 
\citet{dis11} argued that the donor star in the SD model
might shrink rapidly before the WD explosion, because it exhausts
its hydrogen-rich envelope before the SN~Ia explosion during a long
spin-down phase of the rapidly rotating, super-Chandrasekhar mass WD. 
In such a case, the companion star is much smaller
than its Roche lobe, which reduces the shock signature.  
This also explains the lack of hydrogen in the spectra of SNe~Ia
and possibly the unseen companion in the SNR \citep{dis12}.
\citet{hksn12} presented possible evolutionary routes to
super-Chandrasekhar mass WDs in the WD+MS channel of the SD model.
They also compared the spin-down time of WDs with the companion star's
MS lifetime and discussed their final states at explosions, although
their main-focus was to explain the observed extremely luminous SNe Ia.

In this Letter, we apply our method in \citet{hksn12} to WD+RG binaries,
and calculate the WD mass distribution beyond the Chandrasekhar mass limit
for both the WD+MS and WD+RG systems.
We then estimate the brightness distribution of SNe~Ia, assuming that
the brightness depends on the WD mass.  We further confirm that, in most of
the WD+RG systems, the companion has evolved off from a RG to a helium 
(or C+O) WD before the SN~Ia explosion and such a compact companion
does not show any prominent shock signatures nor indications of hydrogen.
In Section \ref{binary_evolution}, we describe our basic assumptions
and methods.  Section \ref{numerical_results} 
presents our numerical results.  Section \ref{discussion} discusses
various perspectives on the progenitors of SNe~Ia.


\begin{figure*}
\epsscale{0.75}
\plotone{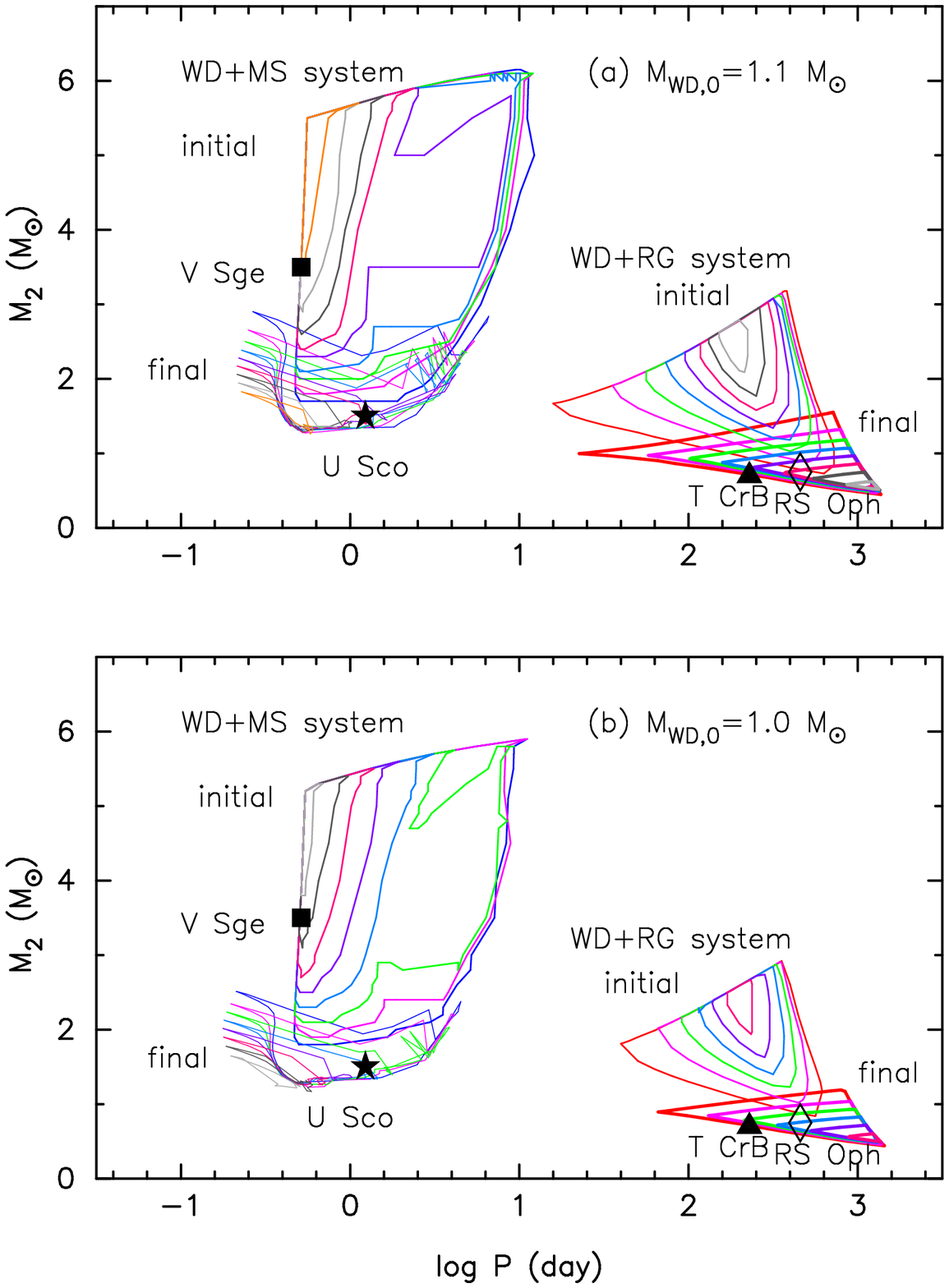}
\caption{
Regions producing SNe~Ia with various initial WD masses
are plotted in the $\log P - M_2$
(orbital period -- donor mass) plane both for the WD+MS system
({\it left}) and the WD+RG system ({\it right}).
The initial system inside the region encircled by a solid line
(labeled ``initial'') is increasing its WD mass up to the 
mass of $M_{\rm WD}= 1.38$, 1.5, 1.6, ..., 2.1, and $2.2~M_\sun$
(from outside to inside) and then reaches the regions labeled 
``final'' when the WD stops growing in mass.
Currently known positions of recurrent novae
and supersoft X-ray sources are indicated by a star mark ($\star$)
for U~Sco \citep[e.g.,][]{hkkm00}, a filled square for V~Sge \citep{hac03kc},
a filled triangle for T~CrB \citep[e.g.,][]{bel98}, and an open diamond
for RS~Oph \citep[e.g.,][]{bra06}.
\label{zregevl_10_11_strip_ms_rg_sc15_2012}}
\end{figure*}


\begin{figure*}
\epsscale{0.75}
\plotone{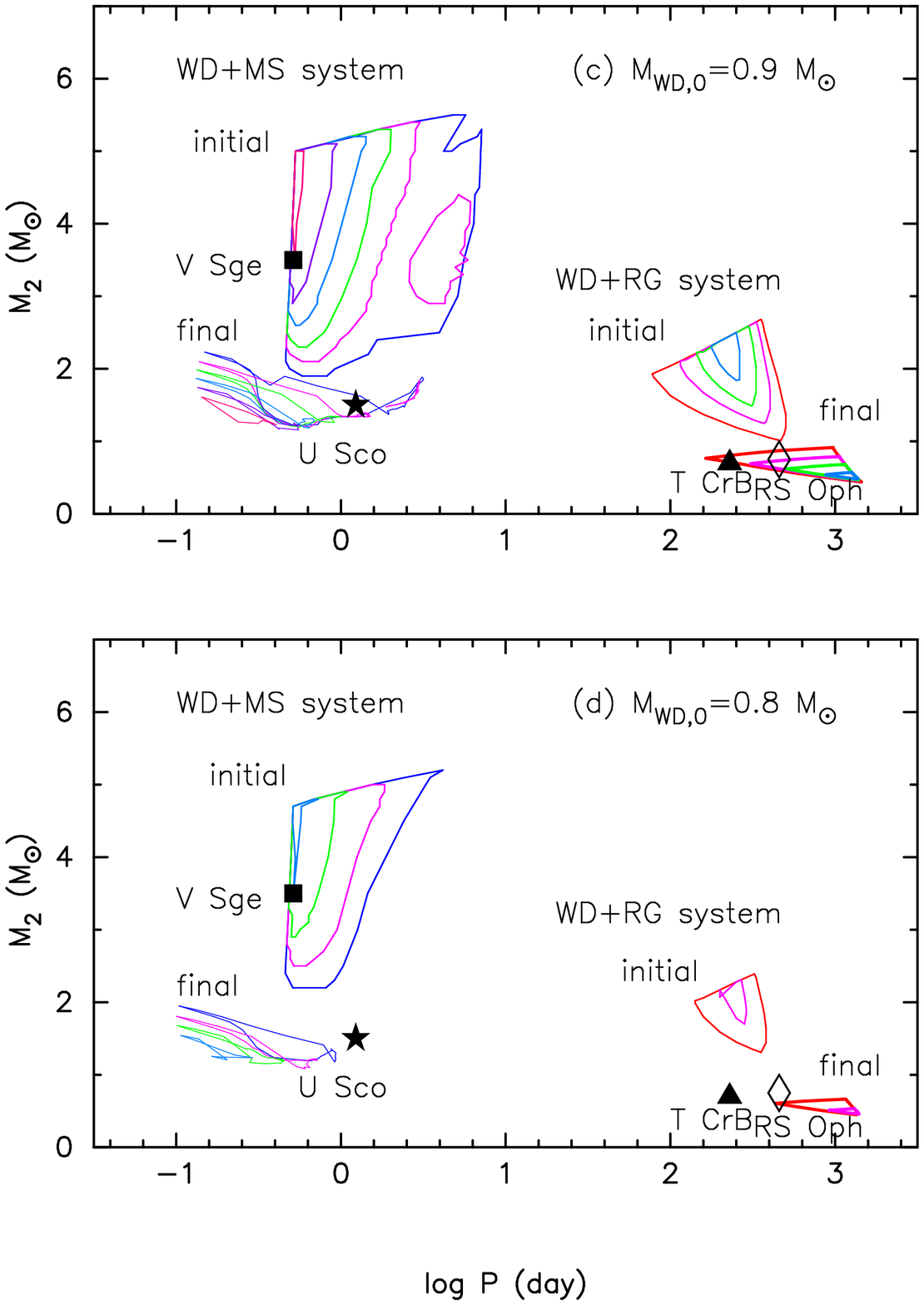}
\caption{
Continued from Figure \ref{zregevl_10_11_strip_ms_rg_sc15_2012}.
\label{zregevl_08_09_strip_ms_rg_sc15_2012}}
\end{figure*}

\section{Binary Evolution beyond the Chandrasekhar mass}
\label{binary_evolution}
Based on the SD model, we followed binary evolutions in which a WD
accretes hydrogen-rich matter from its companion.  There are two well
studied evolutionary paths to SNe~Ia, the WD+MS and WD+RG channels.
Our basic assumptions in binary evolutions are essentially the same as
those in \citet{hkn99, hknu99, hkn08a, hkn08b, hksn12}.
Mass-accreting WDs blow optically thick winds
if the mass transfer rate exceeds the critical rate,
$\dot M_{\rm cr}=6.68\times10^{-7}({{M_{\rm WD}}/{M_\sun}}-0.445)
M_\sun{\rm ~yr}^{-1}$ \citep{hkn96, nom07}.

The WD winds collide with the secondary's surface and strip off its
surface layer.  If the mass-stripping is efficient enough,
the mass transfer rate is attenuated and the binary avoid
the formation of a common envelope even for a rather massive secondary
of $\sim 4$--$6~M_\sun$.  Thus, the mass-stripping effect widens
the donor mass range of SN~Ia progenitors.
We have incorporated the mass-stripping effect
in the same way as in \citet{hkn08a, hksn12}, i.e.,
the mass stripping rate ${\dot M}_{\rm strip}$ is proportional to
the wind mass-loss rate $\dot M_{\rm wind}$ as
${\dot M}_{\rm strip}=c_1{\dot M}_{\rm wind}$.
In this study, we assume $c_1=3$ as a lower representative value for
the WD+MS system, because
\citet{hac03kb, hac03kc} found that $c_1$ is between a few to 10
to reproduce the optical and X-ray light curve behaviors of some
supersoft X-ray sources.  If we adopt a larger value, we could have
a more massive secondary.  For the WD+RG systems,
$c_1$ is calculated from Equation (21) of \citet{hkn99}.

Mass-accreting WDs spin up because of angular momentum gain from
the accreted matter \citep{lan00, uen03, pie03b}.
If the WD rotates rigidly,
its mass can only slightly exceed the Chandrasekhar mass of no rotation, 
$M_{\rm Ch}=1.46(Y_{\rm e}/0.5)^2M_\sun$ with electron mole number 
$Y_{\rm e}$.  If the WD rotates differentially, however, its mass
can significantly exceed $M_{\rm Ch}$ \citep[e.g.,][]{hac86}.
\citet{yoo04} concluded that the WD increases its mass beyond
$M_{\rm Ch}$ when the accretion rate to the WD is as high as
$\dot{M}_{\rm WD}\gtrsim10^{-7}M_\odot$~yr$^{-1}$.
They showed that the gradient of angular velocity is kept around the
critical value for the dynamical shear instability
and that this differential rotation law is strong enough to support
WDs whose masses significantly exceed $M_{\rm Ch}$ \citep{yoo05}.
\citet{pir08} showed that highly differential rotation may not be realized
due to baroclinic instability.  However, his stability condition is not
a sufficient condition but just a necessary condition for instability,
so that his conclusion is premature \citep[see][]{hksn12}.

In the present study, we simply assume that mass-accreting WDs are
supported by the above differential rotation law and the mass can
increase without carbon being fused at the center as long as
$\dot{M}_{\rm WD}>1\times 10^{-7}M_\odot$~yr$^{-1}~(\equiv {\dot M}_{\rm b})$.
We assume that, when ${\dot M}_{\rm WD} < {\dot M}_{\rm b}$,
hydrogen shell-burning occurs intermittently on the WD and the resultant
nova outbursts eject a large part of the envelope mass \citep{hac01kb}.
As a result, the net growth rate of the WD mass is significantly reduced
($\lesssim10^{-8}M_\odot$~yr$^{-1}$).
Then the timescale of angular momentum deposition to the WD core
would become much longer than $\sim10^7$~yr and
comparable to the timescale for the Eddington-Sweet meridional circulation
($\sim10^8$~yr) to redistribute angular momentum.
This redistribution causes a contraction of the WD core and its
central density increases high enough to trigger an SN~Ia explosion
before the WD mass significantly increases.  Thus, if the WD mass
increases beyond $1.38~M_\sun$, we define the WD mass
at the SN~Ia explosion as the WD mass when the mass transfer rate drops
to ${\dot M}_{\rm b}$.


\begin{figure}
\epsscale{1.15}
\plotone{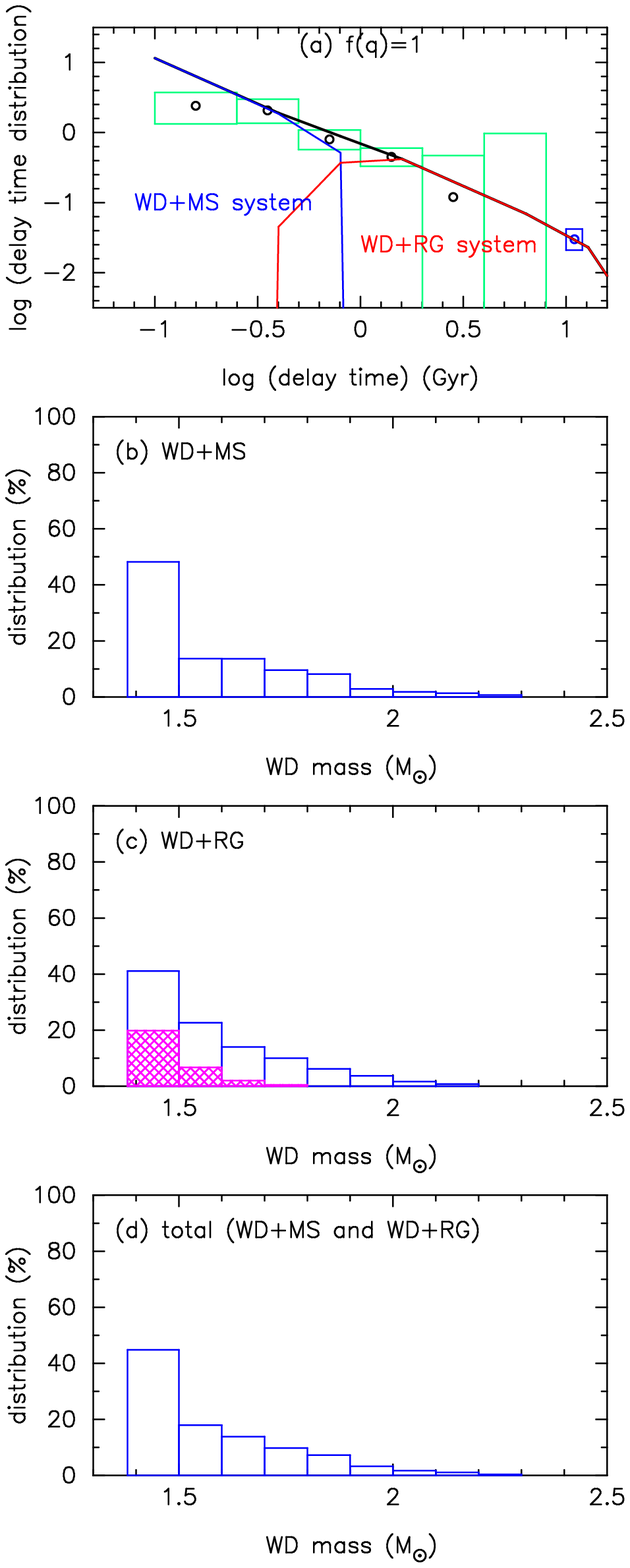}
\caption{
(a) Delay time distribution (DTD) for our WD+MS (blue solid),
WD+RG (red solid) models, and the total (black solid) of them.
The ordinate is the DTD in units of per century and per $10^{10}~L_{K,\sun}$.
Open circles with an open box are the observational DTD taken from
\citet{tot08} (for $< 10$ Gyr) and \citet{man05} (at 11 Gyr).
Each open box indicates a $1~\sigma$ error of each measurement.
(b) Number ratio of the WD masses for the WD+MS systems.
(c) Same as (b), but for the WD+RG systems.
Hatched regions indicate the contribution of 
$M_{2,0} < 1.2~M_\sun$, corresponding to elliptical galaxies.
(d) Same as (b), but for the total of WD+MS and WD+RG systems. 
\label{wd_mass_distribution_c3}}
\end{figure}


\begin{deluxetable}{rrrrr}
\footnotesize
\tablecaption{White dwarf mass distribution of Type Ia supernovae\tablenotemark{a}
\label{wd_mass_distribution_sup_ch}}
\tablewidth{0pt}
\tablehead{
\colhead{WD mass} & \colhead{WD+MS\tablenotemark{b}} 
& \colhead{WD+RG\tablenotemark{c}} & \colhead{ellipticals\tablenotemark{d}} 
& \colhead{total} \\
\colhead{($M_\sun$)} & \colhead{(\%)} & \colhead{(\%)} & \colhead{(\%)}
& \colhead{(\%)}
}
\startdata
1.38--1.5 &  48.2  &   41.1  & 19.8 & 44.8 \\
1.5--1.6 &  13.7  &   22.7  & 6.7 & 17.9 \\
1.6--1.7 &  13.6 &   14.0 & 2.0 & 13.8 \\
1.7--1.8 &  9.6 &  10.0 & 0.5 & 9.8 \\
1.8--1.9 &  8.2 &  6.2 & 0.0 & 7.2 \\
1.9--2.0 &  2.9 &  3.7 & 0.0 & 3.3 \\
2.0--2.1 &  1.8 &  1.6 & 0.0 & 1.7 \\
2.1--2.2 &  1.3 &  0.8 & 0.0 & 1.1 \\
2.2--2.3 &  0.7 &  0.0 & 0.0 & 0.4
\enddata
\tablenotetext{a}{metallicity $Z=0.02$, binary mass ratio distribution
$f(q)=1$}
\tablenotetext{b}{${\nu}_{\rm WD+MS} = 0.0035$~yr$^{-1}$ in our Galaxy}
\tablenotetext{c}{${\nu}_{\rm WD+RG} = 0.0031$~yr$^{-1}$ in our Galaxy}
\tablenotetext{d}{$M_{2,0} < 1.2~M_\sun$ is assumed for elliptical galaxies}
\end{deluxetable}


\begin{deluxetable}{rrrr}
\footnotesize
\tablecaption{WD mass versus maximum luminosity distribution
\label{wd_mass_vs_luminosity_sup_ch}}
\tablewidth{0pt}
\tablehead{
\colhead{WD mass} & \colhead{ratio} & \colhead{$\Delta m_{15}(B)$} 
& \colhead{ratio\tablenotemark{a}} \\
\colhead{($M_\sun$)} & \colhead{(\%)} & \colhead{(mag)} & \colhead{(\%)}
}
\startdata
1.38--1.6 &  62.7  &   1.1--2.1  & 67.4 \\
1.6--1.8 &  23.6 &   1.0--1.1 & 17.3 \\
1.8--2.0 &  10.5 &  0.9--1.0 & 10.2 \\
2.0--2.3 &  3.2 &  0.7--0.9 & 5.1
\enddata
\tablenotetext{a}{taken from \citet{blo12}}
\end{deluxetable}

Figures \ref{zregevl_10_11_strip_ms_rg_sc15_2012} and
\ref{zregevl_08_09_strip_ms_rg_sc15_2012} show
the parameter regions that produce SNe~Ia in the $\log P - M_2$
(orbital period--companion mass) plane for the metallicity of $Z=0.02$.
We plot results for four initial WD masses of $M_{\rm WD,0}=1.1$,
1.0, 0.9, and $0.8~M_\sun$ because (1) no C+O WDs of
$M_{\rm WD,0}\gtrsim 1.1~M_\sun$ are expected
for normal metallicity \citep{hksn12} and
(2) the $0.7~M_\sun$ region is too small for the WD+MS systems
and none for the WD+RG systems.  The WDs inside these SN~Ia regions
(labeled ``initial'') will increase their masses from
(a)~$M_{\rm WD}=1.1$, (b)~1.0, (c)~0.9, and (d)~0.8 $M_\sun$ to 
$M_{\rm WD}=1.38$, 1.5, 1.6, ..., $2.2~M_\sun$ 
($0.1~M_\sun$ step from outside to inside contours)
and reach the regions labeled ``final.''
We stop the binary evolution when the mass transfer rate decreases
to $\dot M_{\rm b}$.  Both the WD+MS and WD+RG systems can produce
a super-Chandrasekhar mass WD up to $\sim2~M_\sun$.

\section{Distribution of WD masses at SN I\lowercase{a} explosion}
\label{numerical_results}
The SN~Ia regions for different initial WD masses,
$M_{\rm WD,0}= 0.7$, 0.8, 0.9, 1.0, and $1.1 ~M_\sun$ are
calculated both for the WD+MS and WD+RG systems.
We then estimate the SN~Ia birth rate in our Galaxy as $\nu_{\rm WD+MS} = 
0.0035$~yr$^{-1}$ and $\nu_{\rm WD+RG} = 0.0031$~yr$^{-1}$ for
the constant star formation rate.
Here we assume the initial distribution of binaries
given by Equation (1) of \citet{ibe84}, i.e.,
$\nu=0.2~\int f(q) d q \int d M/ M^{2.5} \int d \log a 
{\rm ~yr}^{-1}$ and the distribution of mass ratio $f(q)=1$.

We also estimate the delay time distribution (DTD) of SNe~Ia as shown
in Figure \ref{wd_mass_distribution_c3}a,
where the delay time ($t_{\rm delay}$) is the elapsed time from
binary birth to explosion.
The spin-down time of WDs is not included.
The computational method is the same as that in \citet{hkn08b}.
These values are normalized to fit the DTD at 11 Gyr \citep{man05}.
Our DTD shows a featureless power law ($\propto t^{-1}_{\rm delay}$) from 0.1
to 12 Gyr, which is consistent with Totani et al.'s (2008) observation.
The present results are essentially the same as our previous results
by \citet{hkn08b} in which we assume that the WD explodes as an SN~Ia
at $M_{\rm WD}=1.38~M_\sun$.
In general, the WD+MS systems consist of a young population of 
SNe~Ia corresponding to short delay times ($t_{\rm delay} \lesssim 1$~Gyr)
and the WD+RG systems an old population of long delay times
($t_{\rm delay} \gtrsim 1$ Gyr).

We further calculate the distribution of the WD masses at SN~Ia
explosions as in Table \ref{wd_mass_distribution_sup_ch} and
in Figure \ref{wd_mass_distribution_c3}b--d.
We also plot the number ratio of the WDs for the WD+RG systems with
the initial companion masses of $M_{2,0} < 1.2~M_\sun$,
which are expected to occur in elliptical galaxies \citep[see][]{hkn10}.
It is clear that the WD mass distribution in ellipticals is confined
into a narrower range of 1.38--1.6$~M_\sun$ than in late type galaxies. 

\section{Discussion}
\label{discussion}
\subsection{Spin-down time and final fate of WDs}
After the mass transfer rate drops to $\dot M_{\rm b}$, the WD stops
growing in mass.  There are three characteristic mass ranges
for the final evolution of WDs toward an SN~Ia explosion 
\citep[see][for detail]{hksn12}.

(1) In the extremely massive case, the differentially rotating WD
explodes as an SN~Ia soon after the WD mass exceeds $2.4~M_\sun$
owing to a secular instability.  This is not the present case,
because it happens for $M_{\rm WD,0}\gtrsim1.2~M_\sun$ in only
low metallicity environments. 

(2) For the mid-mass range of $M_{\rm WD}=1.5$--$2.4~M_\sun$, the WD
is differentially rotating and its mass exceeds the maximum mass
for rigid rotation.  As angular momentum in the WD core is
lost or redistributed toward rigid rotation, the WD core contracts
until its central density and temperature become high enough
to ignite carbon.  Thus the timescale of contraction
until the SN explosion is $\sim 10^8$ yr due to angular momentum
transport by the Eddington-Sweet meridional circulation.

As for the other angular momentum transport mechanisms,
\citet{ilk11} showed that magneto-dipole radiation leads to
spin-down in a typical timescale of $\sim10^8$--$10^9$~yr for 
$M_{\rm WD}\gtrsim1.6~M_\sun$ when the magnetic field of the WD
is as weak as $\sim10^6$~G.  They also showed that the $r$-mode
instability is not significant in spinning-down WDs.
Thus, we here assume that the Eddington-Sweet meridional circulation
is the most effective process for spin-down.

(3) For the lower mass range of $M_{\rm WD}=1.38$--$1.5~M_\sun$,
the WD can be supported by rigid rotation while it exceeds the critical
mass of non-rotating WDs for carbon ignition.  Thus, the WD contracts
with the spin-down timescale, which is determined by angular momentum loss
from the WD and thus depends on the strength of the
magnetic field of the WD.  The final fate of the WD depends
on the spin-down time as we discuss below.  It may take more than 
$\sim10^9$~yr for weak magnetic fields of $\sim10^6$~G \citep{ilk11}.

If the spin-down time is not much longer than $\sim10^9$~yr, 
the compressional heating due to the spin-down would dominate the
radiative cooling of the WD \citep[see Equation (7) of][]{nom82}.
Because the spin-down time is not unique, its variation causes
a variety of thermal state of WD cores when carbon ignites at
the center.  This would lead to a variation of the carbon ignition
density and thus a variation of $^{56}$Ni mass and brightness
of SNe~Ia even for the same WD mass.

If the spin-down time is much longer than $\sim 10^9$~yr, on the other hand, 
the central density at the carbon ignition could become high enough to
induce collapse \citep{nom91}.  This collapse produces a quite
little amount of $^{56}$Ni as $\sim 10^{-3}M_\sun$ \citep{wana09},
which might correspond to a faint transient.

\subsection{Final fate of WD+MS systems}
In most of the WD+MS systems, the companion remains
to be an MS (central hydrogen burning) star until the
``final'' stage of evolution.  As shown in Figures 
\ref{zregevl_10_11_strip_ms_rg_sc15_2012} and
\ref{zregevl_08_09_strip_ms_rg_sc15_2012}, these systems have an orbital
period shorter than $\sim 1$ day and a companion mass smaller than 
$\sim2~M_\sun$.  This type of pre-supernova binaries satisfy
the constraint on SN~2011fe posed by \citet{liw11}, $M_{\rm MS}<3.5~M_\sun$.

If $M_{\rm WD}=1.5$--$2.4~M_\sun$ and the spin-down time
($\sim10^8$~yr) is shorter than the MS lifetime of the
companion, the WD explodes before the companion evolves off the
main-sequence.  In most of these cases,
the companion's mass further satisfies the condition of 
$M_{\rm MS}\lesssim1~M_\sun$ posed by \citet{bro12} on the SNR~2011fe,
because the companion's mass further decreases 
from the ``final'' mass in Figures 
\ref{zregevl_10_11_strip_ms_rg_sc15_2012} and
\ref{zregevl_08_09_strip_ms_rg_sc15_2012} by the amount of roughly
$\Delta M\sim10^{-8}~M_\sun$~yr$^{-1}\times10^8$~yr$\sim1~M_\sun$,
which is transferred to the WD and ejected by nova outbursts.
On the other hand, if $M_{\rm WD}=1.38$--$1.5~M_\sun$ and 
the spin-down time ($\sim10^9$~yr) is longer than the MS
lifetime of the companion, the companion 
becomes a helium WD and CSM has disappeared.
Such a case might correspond to the case of no CSM like
SN 2011fe \citep{pat11}.

As already discussed in our previous paper \citep{hkn08a},
the mass-stripping effect produces a large amount of CSM,
say, a few to several solar masses
($\approx$ initial mass minus final mass, as seen in Figures
\ref{zregevl_10_11_strip_ms_rg_sc15_2012} and 
\ref{zregevl_08_09_strip_ms_rg_sc15_2012}).
If the spin-down time is short enough, 
or if the WD is forced to be rigidly rotating during accretion
due to strong magnetic fields, the WD may explode in the CSM.
Then we may observe the interaction between the ejecta and
the CSM like in SNe~Ia/IIn.

\subsection{Final fate of WD+RG systems}
In the WD+RG systems, after the mass transfer rate drops to 
$\dot M_{\rm b}$, the companion RG further evolves and finally
becomes a helium (or C+O) WD in a timescale of $< 10^8$~yr.  
This timescale is shorter than the spin-down time in both the
mid-mass range ($1.5~M_\sun<M_{\rm WD}<2.4~M_\sun$: $\sim10^8$~yr for
the Eddington-Sweet circulation) and lower mass range
($1.38~M_\sun<M_{\rm WD}<1.5~M_\sun$: $\sim10^9$~yr for the magneto-dipole
radiation) WDs.  The companion RG has already evolved to a WD
when the primary WD explodes as an SN~Ia.  Therefore, the immediate
progenitor is a wide binary consisting of WD+WD for all the cases.
These immediate progenitors satisfy all the constraints mentioned
in Section \ref{intro}.  It might correspond to SN~2011fe, which shows
no CSM.  It also explains the lack of hydrogen in the spectra of SNe~Ia
and possibly the unseen companions of SN~1572 (Tycho) \citep{ker09},
SN~1006 \citep{ker12}, and SNR~0509-67.5 \citep{sch12}.

In the above discussion, we assumed that the spin-down time is 
$\gtrsim10^8$--$10^9$~yr.  However, if the spin-down time is short enough,
or if the WD is forced to be rigidly rotating during accretion,
the WD may explode before
the companion evolves off the red-giant or asymptotic giant branch.
Then we may observe the interaction between the ejecta and the CSM
like in Kepler's SNR \citep{chi12} and in PTF11kx \citep{dil12}.

\subsection{Variation of Type Ia supernovae}
In our progenitor models, various types of SNe~Ia can be explained 
as follows:  Normal SNe~Ia correspond to the lower mass range of WDs,
 $1.38~M_\sun<M_{\rm WD}\lesssim1.5~M_\sun$ (or $\lesssim1.6~M_\sun$).
The brightness variation can be explained as a variety of spin-down time. 
The brighter group of SNe~Ia such as SN~1991T correspond to the mid-mass range
of WDs, $M_{\rm WD}\gtrsim1.5~M_\sun$ (or $\gtrsim1.6~M_\sun$).
We think that for these brighter SNe~Ia, the brightness variation
stems mainly from a variation of WD mass. 
Here we set the border between the normal and brighter SNe
at about $1.5~M_\sun$.  However, it depends on the timescale of
the Eddington-Sweet circulation which may become
longer near rigid rotation.  Therefore, this border could be as massive
as $M_{\rm WD} \sim 1.6~M_\sun$.

Now we explain the luminosity distribution of SNe~Ia with our model.
In the observation, peak brightnesses of SNe~Ia depend monotonically
on the $\Delta m_{15}(B)$, where $\Delta m_{15}(B)$ is 
the $B$-magnitude decay from the maximum in 15 days.
Table \ref{wd_mass_vs_luminosity_sup_ch} compares
the WD mass distribution in our model with
the observational $\Delta m_{15}(B)$ distribution \citep{blo12}.
For the brighter group of SNe~Ia ($M_{\rm WD}\gtrsim1.6~M_\sun$),
the distribution of $\Delta m_{15}(B)$ is in good agreement with
the WD mass distribution.
This may be a support for our theoretical expectation that the brightness
of SNe~Ia is determined mainly by the WD mass
because the thermal state is similar among various WD cores due to
its relatively short spin-down time.
For the fainter group of SNe~Ia ($M_{\rm WD}\lesssim1.6~M_\sun$),
however, the brightness of SNe~Ia depends not only on the WD mass
but also on the spin-down time as mentioned in the previous subsection.

To summarize, we examined the final fate of the two progenitor models
of SNe~Ia, the WD+MS and WD+RG systems.  A major part of
the WD+MS systems reasonably satisfy the stringent constraints on
SN~2011fe in M101.  Most cases of the WD+RG systems satisfy even 
more stringent constraints on SNR~0509-67.5 posed by \citet{sch12}.

\acknowledgments
This research has been supported in part by the Grants-in-Aid for
Scientific Research of the Japan Society for the Promotion of Science
(22540254, 23224004, 23540262, 24540227) and MEXT (22012003, 23105705)
and by World Premier International Research Center Initiative, MEXT,
Japan.

\end{document}